\documentclass[5p,twocolumn]{elsarticle}

\usepackage{lineno}

\usepackage[percent]{overpic}
\usepackage{etoolbox}
\makeatletter

\usepackage{graphicx,framed,amssymb,mathrsfs,textcomp,comment,fancyhdr,calligra,ifsym,amsmath,amsthm,mathtools,amsmath}

\usepackage[T1]{fontenc}
\usepackage[utf8]{inputenc}
\usepackage{hyperref}
\begin{document}
\title{{\bfseries\large Analytical Benchmarks for Precision Particle Tracking in Electric and Magnetic Rings}}
\author[1,3,4,6]{E.M. Metodiev}
\author[1]{I.M. D'Silva}
\author[1]{M. Fandaros}
\author[4,6]{M. Gaisser}
\author[1,4,5,6]{S. Hac{\i}\"{o}mero\u{g}lu}
\author[1]{D. Huang}
\author[1,3]{K.L. Huang}
\author[1]{\\A. Patil}
\author[1]{R. Prodromou}
\author[1]{O.A. Semertzidis}
\author[1]{D. Sharma}
\author[1]{A.N. Stamatakis}
\author[2]{Y.F. Orlov}
\author[1,4,6]{Y.K. Semertzidis}

\address[1]{Brookhaven National Laboratory, Physics Department, Upton, NY 11973, USA}
\address[2]{Department of Physics, Cornell University, Ithaca, NY, USA}
\address[3]{Harvard College, Harvard University, Cambridge, MA 02138, USA}
\address[4]{Center for Axion and Precision Physics Research, IBS, Daejeon 305-701, Republic of Korea}
\address[5]{Istanbul Technical University, Istanbul 34469, Turkey}
\address[6]{Department of Physics, KAIST, Daejeon 305-701, Republic of Korea}

\renewcommand{\[}{\begin{equation}}
\renewcommand{\]}{\end{equation}}

\date{\today}

\begin{abstract}

A set of analytical benchmarks for tracking programs is required for precision storage ring experiments. To determine the accuracy of precision tracking programs in electric and magnetic rings, a variety of analytical estimates of particle and spin dynamics in the rings were developed and compared to the numerical results of tracking simulations. Initial discrepancies in the comparisons indicated the need for improvement of several of the analytical estimates. As an example, we found that the fourth-order Runge-Kutta/Predictor-Corrector method was slow but accurate, and that it passed all the benchmarks it was tested against, often to the sub-part per billion level. Thus, high precision analytical estimates and tracking programs based on fourth-order Runge-Kutta/Predictor-Corrector integration can be used to benchmark faster tracking programs for accuracy.

\end{abstract}
\maketitle

\section{Introduction}

Analytical estimates for particle dynamics in electric and magnetic rings with and without focusing have been given in a variety of papers and notes. We have used these high precision estimates to provide benchmarks to test the accuracy of precision particle tracking programs.  The term ``focusing'' in this paper is used to  denote ``weak vertical focusing'' unless otherwise indicated.  Horizontal focusing is defined by the vertical focusing plus the geometry of the ring, always conforming with Maxwell's equations.

These benchmarks include:
\begin{itemize}
\item Pitch correction\citep{field, farley1} to particle precession frequency in a uniform B-field with and without focusing.

\item Vertical oscillations and energy oscillations in a uniform B-field with no focusing, electric focusing, and magnetic focusing.

\item Radial and vertical oscillations and energy oscillations in an all-electric ring with and without weak focusing.

\item Synchrotron oscillations and momentum storage with a radio frequency cavity (RF) in a uniform B-field.

\item An EDM signal and systematic error with an RF Wien Filter in a magnetic ring.\end{itemize}

In the analytical estimates that follow, we define $\gamma_0$ as the Lorentz factor of the design particle in the ring. The vertical pitch angle $\theta_y$ of a particle is defined such that $\theta_y = \beta_z/\beta_\theta$ where $(\beta_\rho,\beta_\theta,\beta_z)= \vec v/c$ in cylindrical coordinates. The field focusing index is $n$, with $n = -(dB/B_0) / (dr/r_0)$ a number with range  $0<n<1$.

These analytical estimates provide a means of benchmarking particle tracking programs. A precision tracking method that successfully passes all benchmarks can provide a baseline to benchmark faster programs. We study the developed benchmarks in the context of Runge-Kutta/Predictor-Corrector integration. The validity of the developed benchmarks extends to any precision tracking program.

\section{Motivation}

Precision experiments such as the Muon ($g-2$) and Storage Ring Electric Dipole Moment (EDM) experiments\citep{muong2,pedm,pedm2} require measurements of sub-part per million (ppm) accuracy. In the case of a proton or deuteron Storage Ring EDM experiment, a tracking program of extraordinary precision is required to estimate the spin coherence time of the particle distribution and various lattice parameters, as well as to estimate the values of systematic errors associated with the experiment. Many commonly used beam and spin dynamics programs ignore, or erroneously account for, second and higher-order effects. Tracking in an electric storage ring poses the additional challenge of conforming with total-energy conservation while accounting for higher-order effects.

Numerical integration with a sufficiently small step size allowed to run for a sufficiently long time may reproduce the analytical result with high accuracy. A tracking program to be used for estimates and investigations in precision experiments must be optimized to be as fast and accurate as possible. This calls for a well-tested and robust procedure to benchmark the accuracy of tracking programs in situations relevant to the experiments.

We summarize and derive analytical solutions to the equations of motion of a particle in various electric and magnetic rings. In several cases, comparison of the analytical estimate with precision tracking results identified discrepancies and indicated the need to improve the estimates.  In this way it was determined that the total correction due to vertical particle oscillations, the so-called pitch effect, can be significantly reduced\citep{jparc}. These analytical estimates provide individual benchmarks for tracking programs. A program well-benchmarked against these estimates would provide additional, more flexible means to benchmark faster programs in different conditions. We benchmarked a program based on Runge-Kutta/Predictor-Corrector method\citep{hamming} against the developed analytical estimates.

Runge-Kutta/Predictor-Corrector integration should reproduce the analytical estimates to sub-ppm accuracy on a time scale on the order of hours, in order to be a feasible candidate for benchmarking faster programs. Together, the analytical estimates and a program based on Runge-Kutta/Predictor-Corrector integration provide a powerful tool for benchmarking precision tracking programs for Muon $(g-2)$, Storage Ring EDM, and other precision physics experiments requiring high precision beam and spin dynamics simulation.

\section{Precision Tracking}
For a particle of mass $m$ and charge $e$, there are two differential equations that govern particle and spin dynamics. For particle velocity $\vec\beta$ and rest spin $\vec s$ in external fields, the equations are\citep{jackson}:
\begin{equation}\frac{d\vec\beta}{dt} = \frac{e}{m\gamma c}\left[\vec E + c\vec \beta\times\vec B - \vec \beta(\vec\beta\cdot\vec E)\right],\end{equation}
and the T-BMT equation, with an anomalous magnetic moment $a$ of the particle:
\begin{align}\frac{d\vec s}{dt}& = \frac{e}{m}\vec s\times \left[ \left(a + \frac{1}{\gamma}\right)\vec B\nonumber \right. \\ & \left. - \frac{a\gamma}{\gamma + 1}\vec \beta(\vec \beta \cdot\vec B) -\left(a + \frac{1}{\gamma + 1}\right)\frac{\vec\beta\times\vec E}{c}\right].\end{align}

The Predictor-Corrector integration was used with a step size of $1-10$ps to numerically solve the two differential equations with the corresponding initial conditions. Although the method is relatively slow, it is very simple and accurate.  This method uses the Runge-Kutta method to start the integration process and we refer to it as Runge-Kutta/Predictor-Corrector in this document.

\section{Magnetic Ring}
A magnetic ring consists of a uniform magnetic field $\vec B$, taken to be in the vertical direction. The correction $C$ to the precession frequency due to a vertical pitch is defined by $\omega_m = \omega_a(1 - C)$, where $\omega_a$ is the ($g-2$) frequency\citep{bennett1} for a particle with anomalous magnetic moment $a$. The predicted correction is\citep{farley1}:
\[C = \frac14\theta_0^2\{1 - (\omega_a^2 + 2a\gamma^2 \omega_p^2)/\gamma^2(\omega_a^2 - \omega_p^2) \}. \label{mrnf1}\]
with $\omega_p=2 \pi f_p$, where $f_p$ is the vertical (pitch) oscillation frequency.

\subsection{No Focusing}
When there is no focusing or when $\omega_p  \ll \omega_a$, the correction from Equation \ref{mrnf1} becomes:

\[C = \frac14\beta^2\theta_0^2,\label{mrnf2}\]
where for linear oscillations, $\langle \theta_y^2 \rangle = \frac12 \theta_0^2$, where $\theta_0$ is the maximum pitch angle of the particle trajectory.

\begin{figure}[H]
\centering
\scalebox{.65}[.65]{\includegraphics{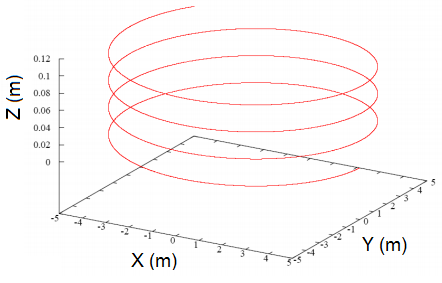}}
\caption{\label{pitch1}The particle path in Cartesian coordinates in a uniform B-field with pitch angle $\theta_y = 1.0$mrad, for a ring with a 5m radius.}
\end{figure}

For a particle with $\beta = 0.972$ and a  constant 1.0mrad vertical pitch as shown in Figure \ref{pitch1}, the simulated correction to the ($g-2$) precession frequency of $0.2361$ppm is in very good agreement with the analytically predicted value of $0.2363$ppm using Equation \ref{mrnf2}.

Checking over several values of $\theta_y$ confirms that the analytic expression and the pitch correction in the tracking simulation agree for small $\theta_y$, as expected.

\subsection{Weak Magnetic Focusing}
When there is magnetic focusing and when $\omega_p  \gg  \omega_a$, the correction from Equation~\ref{mrnf1} becomes:

\[C = \frac14\theta_0^2(1+2a).\label{mrnf3}\]

The analytical estimate\citep{cs} for the average particle radial deviation from the ideal orbit with radius $r_0$, with weak magnetic focusing index $n$, takes the form:
\[\left< \frac{\Delta r}{r_0} \right>= \alpha_p\left<\frac{\Delta p}{p_0}\right>= -\frac{1}{1 - n}\langle \theta_y^2 \rangle\label{mrmf1},\]
for a vertical pitch frequency significantly greater than the ($g-2$) precession frequency of the particle, where $\alpha_p$ is the momentum compaction factor.

\begin{figure}[H]
\centering
\scalebox{.29}[.29]{\includegraphics{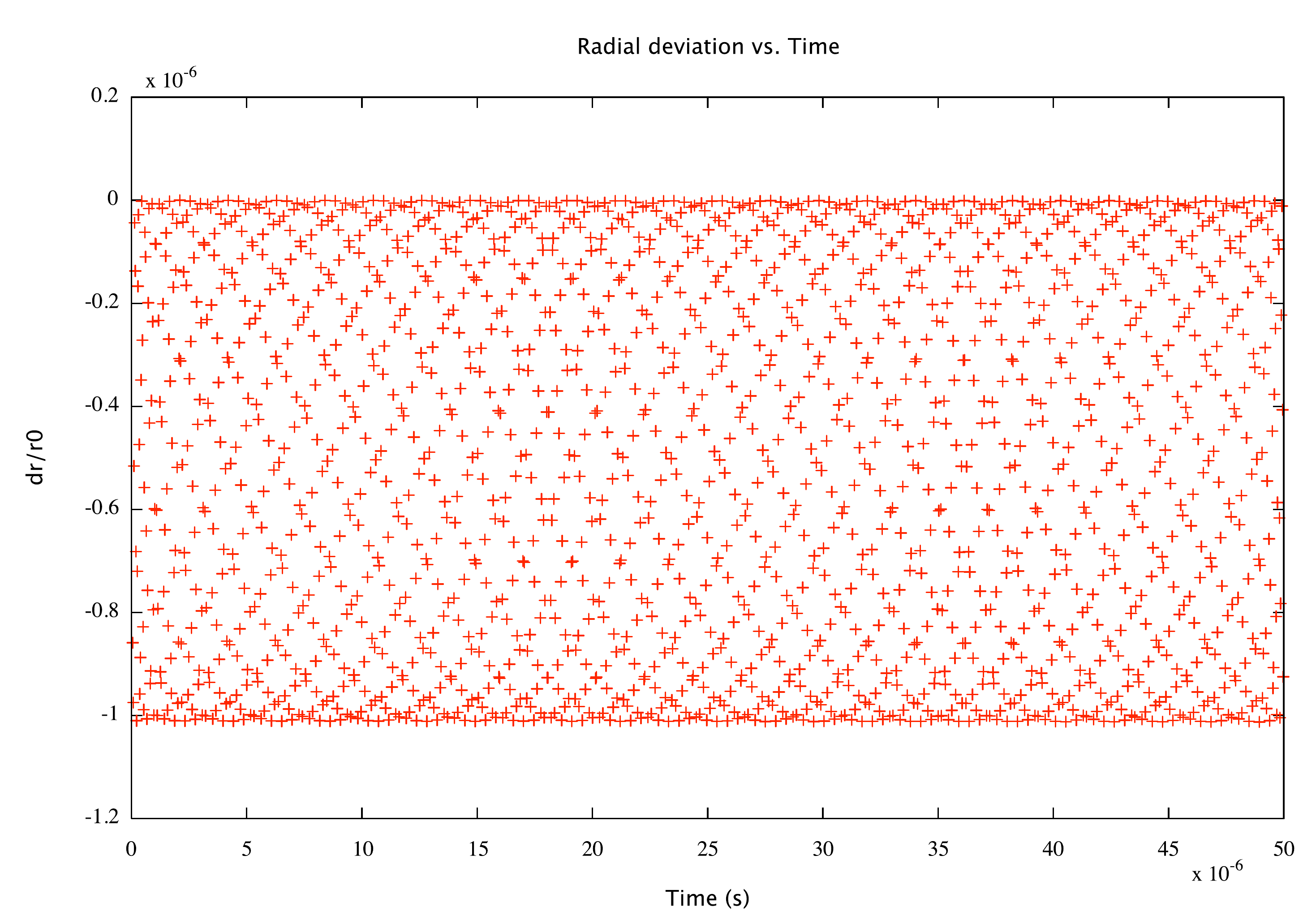}}
\caption{\label{drmrmf}The particle deviation from the ideal radial position over time, modulo 50$\mu$s. The simulations used a maximum pitch angle of $\theta_0 = 1$mrad and magnetic focusing with field index $n = 0.01$.}
\end{figure}

\begin{figure}[H]
\centering
\scalebox{1.25}[1.25]{\includegraphics{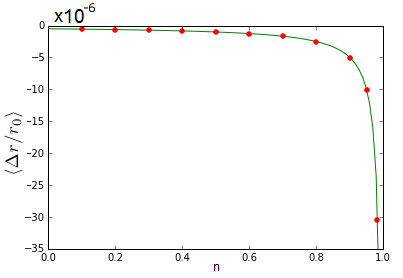}}
\caption{\label{bfdr}The average $\langle\Delta r/r_0\rangle$ versus the field focusing index $n$. The solid line represents the predicted values while the points are the results of tracking. The simulation used a maximum pitch angle of $\theta_0 = 1.0$mrad.}
\end{figure}

Equation \ref{mrmf1} predicts an average radial deviation $\langle\Delta r/r_0\rangle$ of $-5\times 10^{-7}$ using $\theta_0 =$ 1mrad and a field index $n = 0.01$, consistent with the tracking results shown in Figure \ref{drmrmf} to sub-part per billion (ppb) level. The dependence of $\left< \Delta r/r_0\right>$ on the field index is shown to hold over a range of $n$ values in Figure \ref{bfdr}.

In a  continuous storage ring with weak focusing, field strength $B_0$,  and ring radius $r_0$,  the vertical and horizontal magnetic field components around the ideal trajectory can be expressed to second-order in the vertical position $y$ as:
\[B_x(x,y) = -n\frac{B_0}{r_0}y\]\[B_y(x,y) = B_0 - n\frac{B_0}{r_0} x + n \frac{B_0}{r_0}\frac{y^2}{2r_0},\]
where the nonlinearity arises from the application of Maxwell's equations in cylindrical coordinates.  The horizontal and vertical tunes are given by $\nu_x = \sqrt{1 - n}$ and $\nu_y = \sqrt{n}$ respectively.

We make use of the relations\citep{cs} from Equation \ref{mrmf1}:
\[\frac {\left< x \right >} {r_0} = -\alpha_p \frac {\theta_0^2} {2} = -\frac {1} {1-n} \frac {\theta_0^2} {2}, \]
and $\theta_0=y_0/(r_0/\sqrt{n})$, where $y_0$ is the maximum vertical excursion. From this we see that:
\[ C_B = \left<  \frac {B_y} {B_0} -1 \right > = -\frac {n} {r_0} \left < x \right > + n \frac {y_0^2} {4 r_0^2} = \frac {n} {1-n} \frac {\theta_0^2} {2} + \frac {\theta_0^2} {4}.\]

By considering the time-averaged relative B-field change, calling $C_B$ the modification due to the different B-field encountered by the particle, we find that the correction $C^* = C - C_B$ to the ($g-2$) precession frequency in a magnetic storage ring with weak focusing, i.e. 
$\omega_m = \omega_a(1 - C^*)$, is given by the expression:
\[C^* = -\left(\frac{n}{1 - n}-a\right) \frac {\theta_0^2} {2},\label{mrmf2}\]
where we see that several terms of the inhomogeneous B-field correction and the correction in Equation \ref{mrnf3} cancel, leaving a small correction. The necessity of including the second-order inhomogeneous magnetic field contributions was overlooked by previous authors. Investigations with precision tracking identified the discrepancy and motivated the improvement of the analytical estimate.

This holds for a vertical pitch frequency much greater than the ($g-2$) precession frequency of the particle, which for a weak focusing ring means $\sqrt{n} \gg a \gamma$.   Equation \ref{mrmf2} implies that the pitch effect can, in principle, be made to vanish for $n=a/(1+a)$, but the condition $\sqrt{n} \gg a \gamma$ makes it rather difficult to achieve.  To test the tracking program we introduced a particle with 10 times the muon mass, with same $a$ value as the muon, stored in a ring radius of 7.112m.  The program indeed showed that the pitch correction vanishes with an uncertainty at the part per billion (ppb) level when $n = a/(1 + a)$ was used.

\begin{figure}[H]
\centering
\scalebox{.3}[.3]{\includegraphics{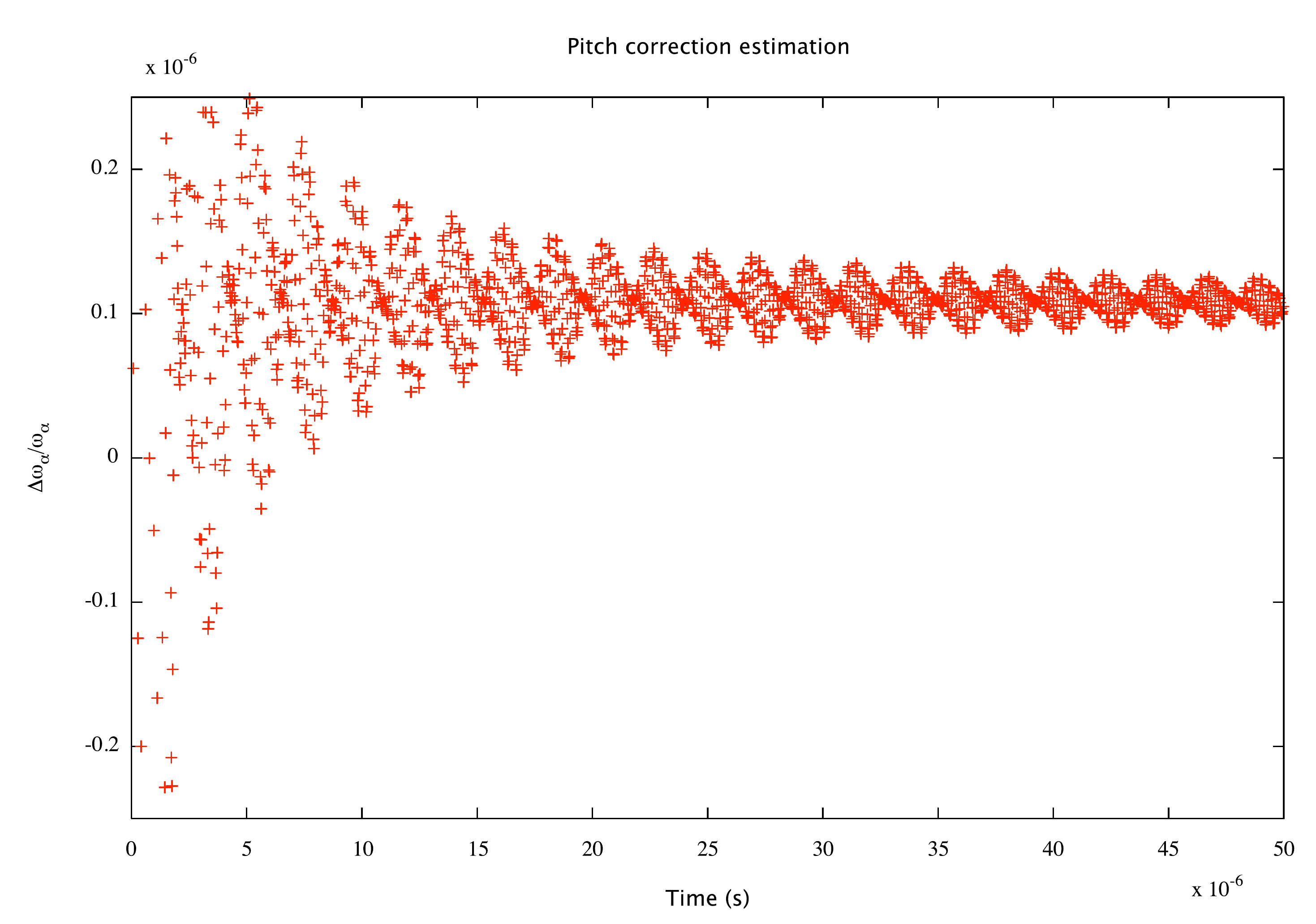}}
\caption{\label{pitch2a}The pitch correction to the ($g-2$) frequency with angle $\theta_0 = 1.0$mrad, $\gamma = 29.3$, and $n=0.18$, is 0.109ppm, consistent with Equation \ref{mrmf2} to sub-ppb level. }
\end{figure}

For realistic muon parameters, the observed ($g-2$) frequency is off from its correct value by +0.109ppm, for a vertical maximum pitch angle $\theta_0=1$mrad, and $n=0.18$, consistent with the offset shown in Figure \ref{pitch2a} to sub-ppb level.  

\begin{figure}[H]
\centering
\scalebox{0.68}[0.68]{\includegraphics{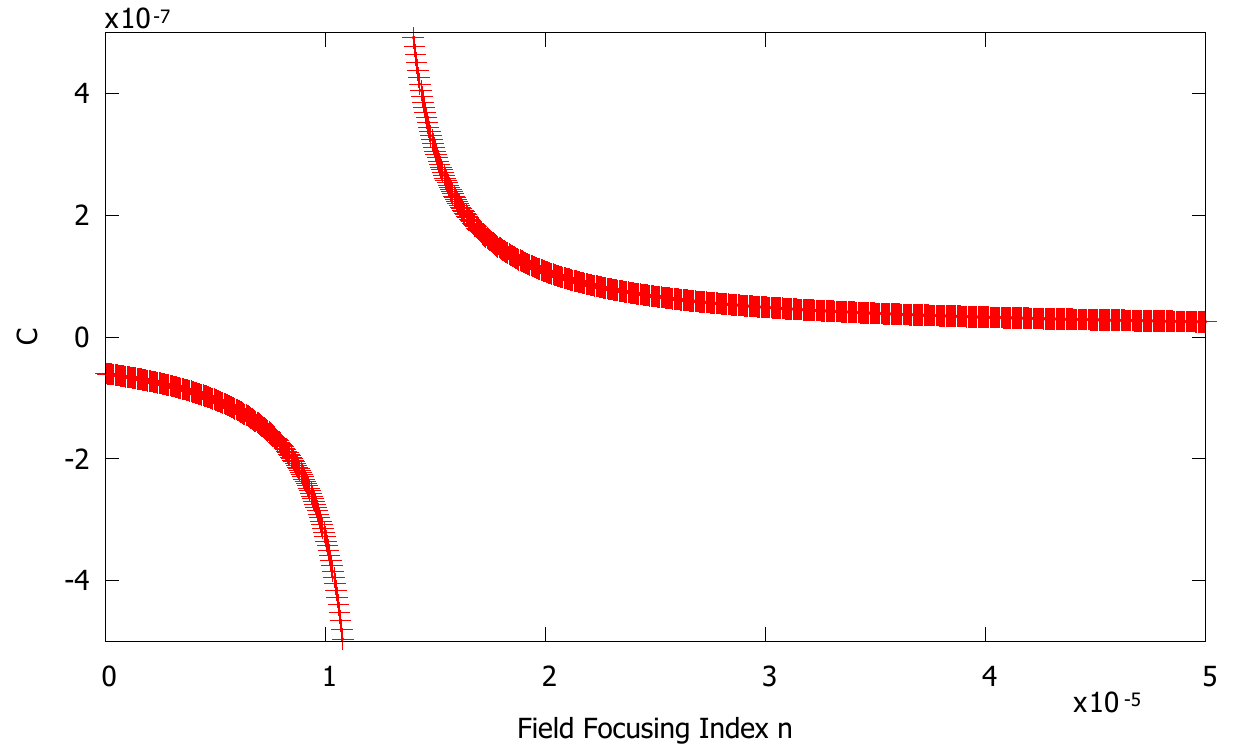}}
\caption{\label{pitch2}Parameter $C$ from Equation \ref{mrnf1} for the pitch correction to the ($g-2$) frequency over a range of $n$ values using $\theta_y = 1.5$mrad and $\beta = 0.94$. A resonance occurs when  $\omega_p = \omega_a$, at $n\approx 1.25\times 10^{-5}$, as expected. }
\end{figure}

A resonance of the pitch effect correction occurs when the vertical pitch frequency $\omega_p$ is equal to the ($g-2$) precession of the stored particles $\omega_a$. The correction $C$ approaches Equation \ref{mrnf2} for $\omega_p \ll \omega_a$ and Equation \ref{mrnf3} for $\omega_p \gg \omega_a$. The full range of pitch corrections $C$ over a range of index values is shown in Figure \ref{pitch2}. When all the fields are taken properly into account, as shown above, the tracking results reproduce the same curve to sub-ppb level for $\Delta \omega_a/\omega_a$.

A comparison of the frequency shift predicted by Equation \ref{mrmf2} and the results from tracking is given in Table \ref{mrmft1}. The analytical estimates of the pitch correction and the tracking results are in very good agreement, better than ppb level.  This level of precision is adequate for the Muon $g-2$ experiments currently underway~\cite{jparc,fnal_g2}, both aiming for better than 0.1ppm total systematic error.
\begin{table}[H]
\caption{Comparison of the frequency shift estimated using Equation \ref{mrmf2}  and the tracking results.  The tracking results assume a muon with $\gamma = 29.3$, stored in a magnetic ring with magnetic focusing and a radius $r_0 = 7.112$m.  The vertical angle used is $\theta_y = 0.5$mrad. The observed $g-2$ frequency is shifted higher by the small factors given below depending on the $n$-value used.}
\vspace{1mm}
\centering
\begin{tabular}{c c c} 
\hline\hline 
$n$ & Estimation (ppb) & Tracking (ppb) \\ [0.5ex] 
\hline 
0.01  & 1.1 & 1.0 \\ 
0.02  & 2.4 & 2.4 \\
0.03  & 3.7 & 3.6 \\
0.05  & 6.4 & 6.4 \\
0.08  & 10.7 & 10.8 \\
0.10  & 13.7 & 13.7 \\
0.137  & 19.7 & 19.9 \\
0.237  & 38.7 & 38.8 \\ [1ex]
\hline 
\end{tabular}
\label{mrmft1}
\end{table}

\subsection{Weak Electric Focusing}
In the case of electric focusing in a uniform magnetic ring, the expected precession frequency correction due to the pitch effect is~\cite{farley1}:
\[C = \frac14\theta_0^2\{\beta^2 - (a^2 \beta^4 \gamma^2 \omega_p^2)/(\omega_a^2 - \omega_p^2) \}. \]

and
\[C =\frac14\theta_0^2 \beta^2 (1+a),\]
for $\omega_p \gg \omega_a$ and for the particle at the magic momentum\citep{pedm,pedm2} such that an electric field does not affect the ($g-2$) precession.

\begin{figure}[H]
\centering
\scalebox{.3}[.3]{\includegraphics{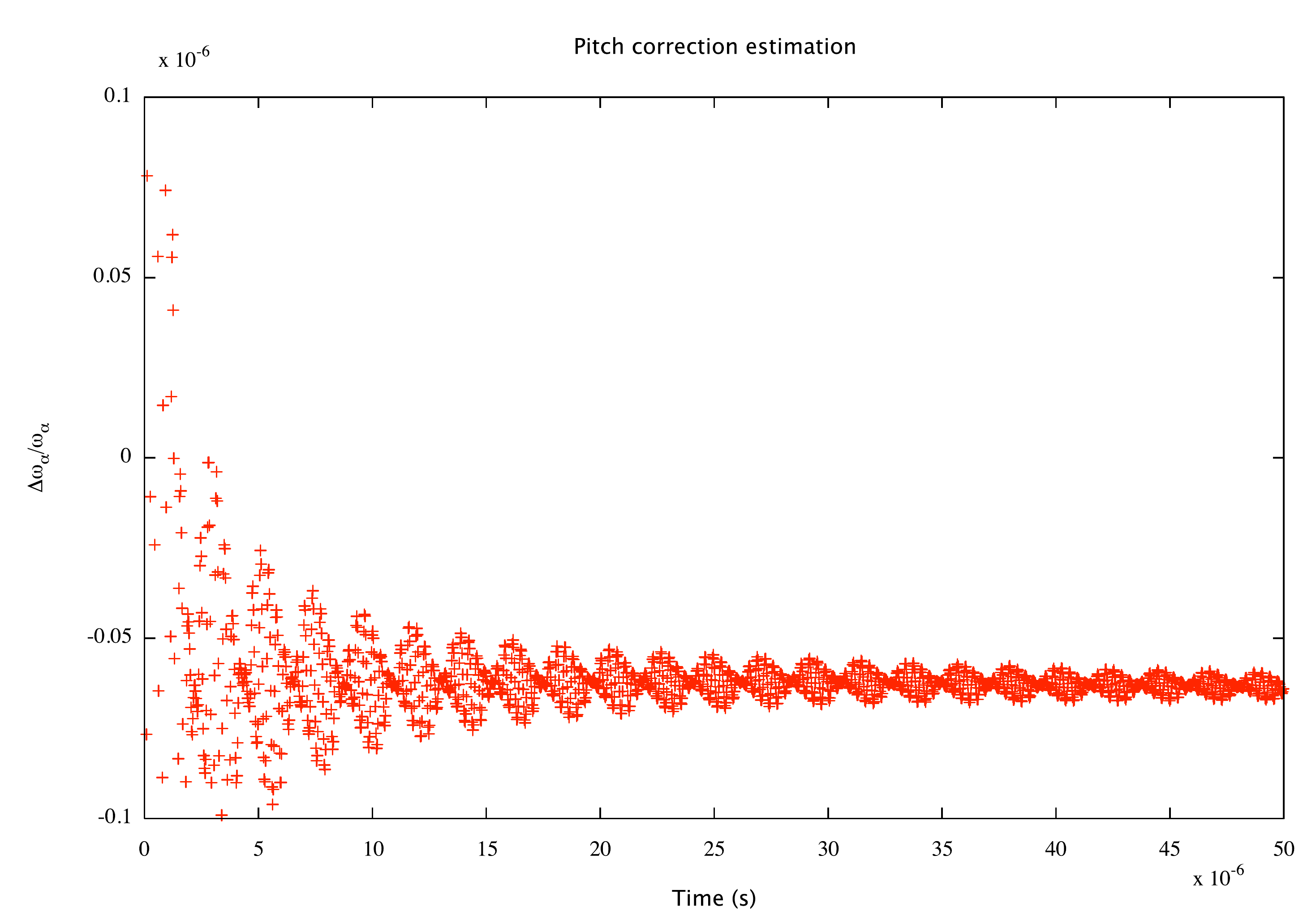}}
\caption{\label{efocus}The ($g-2$) relative difference estimated from tracking using electric focusing in a magnetic ring with a maximum pitch angle of $\theta_0 = 0.5$mrad. To obtain the correct $g-2$ frequency, the correction $0.25 \times \theta_0^2$ needs to be added to the observed frequency.  Again, the tracking results are consistent to sub-ppb level with the predictions.}
\end{figure}

For a maximum vertical pitch of $\theta_0 = 0.5$mrad, the analytically estimated pitch correction of $0.0624$ppm for magic momentum muons is very close to the result from tracking, shown in Figure \ref{efocus}, consistent to sub-ppb level.

The results from tracking match the predicted value to sub-ppb accuracy. Thus, we conclude that the analytical estimates and the Runge-Kutta/Predictor-Corrector integration method have passed the magnetic ring tracking benchmarks.

\subsection{Radio Frequency Cavity}

The synchrotron oscillation frequency $f_s$ of a particle in a uniform B-field with an RF is:
\[f_s = Q_s f_c,\]
where $f_c$ is the cyclotron frequency and $Q_s$ is the synchrotron tune, which satisfies:
\[Q_s^2 = \frac{e V_0 \eta_c h}{2\pi c p_0 \beta^2}.\]

In the expression above, $e$ is the elementary charge, $V_0$ is the voltage of the RF cavity, $h$ is the harmonic of the RF cavity used, and $p_0$ is the ideal momentum. The value of $\eta_c$, the so-called slip factor, is determined from the expression:
\[\eta_c = \alpha_p - \frac{1}{\gamma_0^2} = \frac{1}{1 - n} - \frac{1}{\gamma_0^2}.\]

\begin{figure}[H]
\centering
\scalebox{1.25}[1.25]{\includegraphics{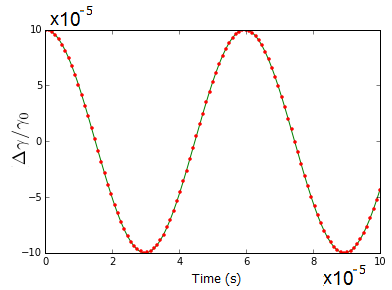}}
\scalebox{1.25}[1.25]{\includegraphics{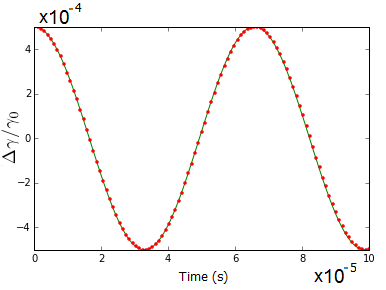}}
\caption{\label{rfSO}The synchrotron oscillations of a particle in a uniform B-field with an RF with (top) $n = 0.18$ and (bottom) no vertical focusing. The solid line represents the estimated oscillations; the points are the results of tracking. }
\end{figure}

Using a particle with charge $e$, $p_0 = 3.094$GeV/c, and $\gamma_0 = 29.3$, and using a 20cm RF cavity, with $V_0 = 100$kV and harmonic $h = 1$, the predicted synchrotron frequency with $n = 0.18$ is $f_s = 16.8$kHz and with no vertical focusing it is $f_s = 15.2$kHz. Comparing these calculations with the results of tracking in Figure \ref{rfSO} shows close agreement between the tracking simulation and the estimation, verified at the 0.1\% level.

The maximum momentum storage range of the RF cavity\cite{rfr} is given by the expression:
\[\left(\frac{\Delta p}{p_0}\right)_{max} = \sqrt{\left|\frac{2eV_0}{\pi h \eta_c \beta c p_0}\right|},\label{rfstor}\]
around the ideal particle with momentum $p_0$.

\begin{figure}[H]
\centering
\scalebox{0.402}[0.402]{\includegraphics{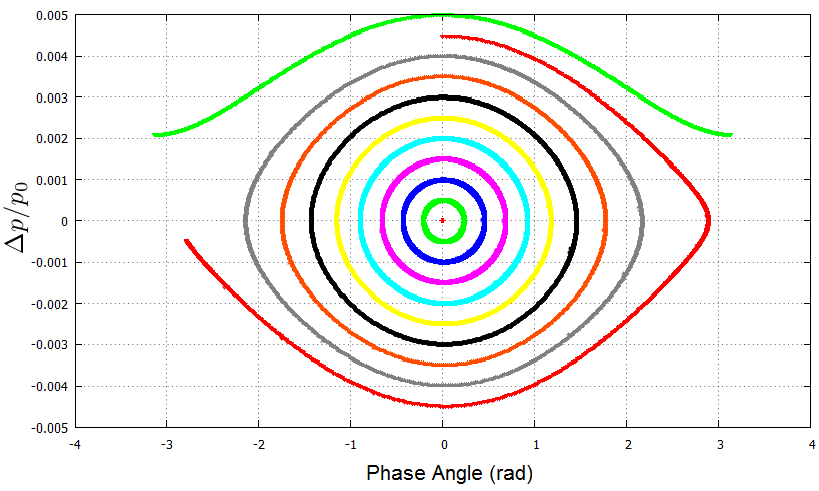}}
\caption{\label{rfPD} A phase diagram of particle momenta in a uniform B-field with an RF and no vertical focusing. The closed energy oscillations around the synchronous particle define the storage region of the RF. The unstored particle is close to the boundary of the storage region.}
\end{figure}

Using the above RF parameters and particle values, the maximum stored momentum is estimated from Equation \ref{rfstor} to be $(\Delta p/p_0)_{max} = 0.00454$ with no vertical focusing. The RF phase diagram shown in Figure \ref{rfPD} is consistent with the estimated value and illustrates the momentum storage region of the configuration.

The analytical estimates for the maximum stored momentum and the synchrotron frequency were matched closely by the Runge-Kutta/Predictor-Corrector tracking method. Thus, we see agreement to the desired accuracy between the analytical estimates and results of Runge-Kutta/Predictor-Corrector integration in the case of a magnetic ring.

\section{Electrostatic Ring}

In cylindrical coordinates, the electric field with an index $m$ power law dependence on radius at $y = 0$ is:
\[\vec E(r,0) = E_0\frac{r_0^{1 + m}}{r^{1 + m}}\hat r,\]
where $\hat y$ is the vertical direction and $\hat r$ is in the radial direction.

In a uniform all-electric ring, we have found that the $y \rightarrow - y$ and rotational symmetries allow the radial and vertical electric field components to be found exactly:
\begin{equation} E_y(r,y) = E_0\frac{r_0^{1 + m}}{r^{1 + m}}\frac{my}{r} \,_{2}F_1\big(1 + \frac{m}{2},1+ \frac{m}{2}; \frac32; - \frac{y^2}{r^2}\big)\label{eqey} \end{equation}
\begin{equation} E_r(r,y) = E_0\frac{r_0^{1 + m}}{r^{1 + m}}\,_{2}F_1\big(1 + \frac{m}{2}, \frac{m}{2};\frac12; - \frac{y^2}{r^2}\big)\label{eqer}, \end{equation}
for all $m>0$, where $_{2}F_1$ is the ordinary hypergeometric function. 
The field index is $n = m + 1$, and $m = 0$ corresponds to cylindrical plates with no vertical focusing, $m = 1$ corresponds to spherical plates, and so on. The focusing $m$ value depends on the choice of electrode profile.

For $m = 0$, the electric field is that of a uniform cylindrical capacitor. The fields were taken to fifth-order in $y/r$ when implemented in the tracking program. The expansion to second-order is shown below:

\begin{equation}E_r(r,y) = E_0\frac{r_0^n}{r^n}\left[1 - \frac12 (n^2 - 1)\frac{y^2}{r^2} + \mathscr O\left(\frac{y^4}{r^4}\right)\right]\label{yaneq}\end{equation}
\begin{equation}E_y(r,y) = E_0\frac{r_0^n}{r^n}\left[ (n-1)\frac{y}{r} + \mathscr O\left(\frac{y^3}{r^3}\right) \right].\end{equation}

 The contributions of the higher-order electric field terms were found to be negligible for tracking. The second-order term is significant for the analytical estimates. The fields given in Equations \ref{eqey} and \ref{eqer} describe the field configuration considered for an electric ring.

In an all-electric ring, the kinetic energy changes with the radial position, which provides additional horizontal focusing. The horizontal and vertical tunes\citep{laslett,satish} are given by $\nu_x = \sqrt{1 - m + 1/\gamma^2}$ and $\nu_y = \sqrt{m}$ respectively, in an electric ring with weak focusing.

\subsection{No Focusing, including an RF cavity}

With no focusing in the ring, we have $n = 1$ and thus $m = 0$, corresponding to concentric cylindrical plates.  We also include an RF-cavity, which fixes the particle revolution frequency.  Y. Orlov\citep{orlov1, orlov2} and I. Koop\citep{mane1} solved the orbital motion for an electrostatic field with no focusing. In this case, the estimates for the average values of $\Delta\gamma/\gamma_0$ and $\Delta r/r_0$ take the following form:
\[\left< \frac{\Delta\gamma}{\gamma_0} \right>= \langle \theta_y^2 \rangle \frac{\gamma_0^2 - 1}{\gamma_0^2 + 1},\label{ernf1}\]
\[\left< \frac{\Delta r}{r_0} \right>= -\frac{\langle \theta_y^2 \rangle}{2} \frac{\gamma_0^2 - 1}{\gamma_0^2 + 1}.\label{ernf2}\]

Note that these values depend only on the particle ideal Lorentz factor $\gamma_0$ and the pitch angle, not on the ring radius, plate spacing, or electric field strength.

\begin{figure}[H]
\centering
\scalebox{1.25}[1.25]{\includegraphics{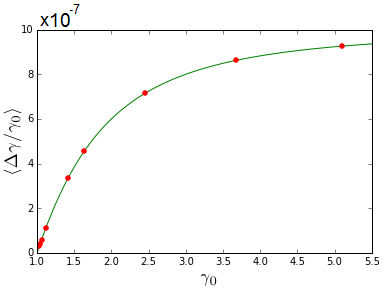}}
\scalebox{1.25}[1.25]{\includegraphics{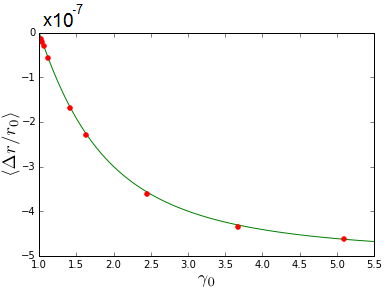}}
\caption{\label{drernf}The average  $\langle\Delta\gamma/\gamma_0\rangle$ (on top) and average $\langle\Delta r/r_0\rangle$ (on bottom) versus the ideal $\gamma_0$ of the proton. The solid lines represent the predicted values; the points are the results of tracking. The simulation used a pitch angle of $\theta_y = 1.0$mrad, an RF-cavity and no vertical focusing.}
\end{figure}

The precision tracking results for the two parameters and the predicted values of Equation \ref{ernf1} and Equation \ref{ernf2} are shown in Figure \ref{drernf}. We see close agreement between the expected value and the values calculated through tracking.  Incidentally we found from tracking that, without including an RF-cavity, $\left< \frac{\Delta r}{r_0} \right>$ from Equation \ref{ernf2} becomes:
\[\left< \frac{\Delta r}{r_0} \right>= -{\langle \theta_y^2 \rangle} \frac{\gamma_0^2 }{\gamma_0^2 + 1},\label{ernf3}\]
whereas Equation \ref{ernf1} remains the same.

\subsection{Weak Electric Focusing, including an RF cavity}
With weak focusing such that $0 < m \ll 1$, the parameters analytically estimated by Y. Orlov\citep{orlov1, orlov2} are given by Equations \ref{eqq} and \ref{eqq2} below:

\[\left< \frac{\Delta\gamma}{\gamma_0} \right>= 0,\label{eqq}\]
\[\left< \frac{\Delta r}{r_0} \right>= -\frac12 \langle \theta_y^2 \rangle,\label{eqq2}\]
which hold for times much larger than the period of vertical oscillations. Note that these values depend only on the pitch angle and not on the ring geometry, ideal $\gamma_0$ or the field focusing index.

There is an apparent gap between Equations \ref{ernf1} and \ref{ernf2} and Equations \ref{eqq} and \ref{eqq2} in the limit as $m\rightarrow 0$. The transition between focusing and no focusing can exist since the latter formulas hold only for averages over times much larger than the period of vertical oscillations\citep{orlov2}.

\begin{figure}[H]
\centering
\scalebox{1.25}[1.25]{\includegraphics{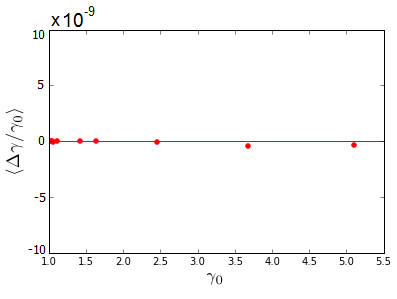}}
\caption{\label{mrwf1}The average $\langle\Delta\gamma/\gamma_0\rangle$ versus the ideal $\gamma_0$ of the proton over a variety of focusing $n$ values. The solid line represents the predicted values; the points are the results of tracking. The simulation used $\theta_0 = 1.0$mrad, an RF-cavity and vertical focusing.}
\end{figure}

\begin{figure}[H]
\centering
\scalebox{1.25}[1.25]{\includegraphics{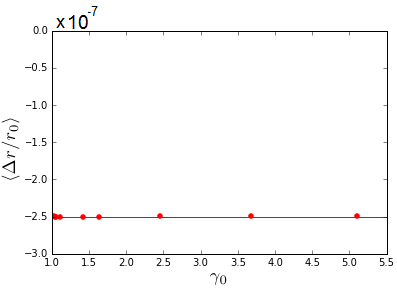}}
\caption{\label{mrwf2}The average $\langle\Delta r/r_0\rangle$ versus the ideal $\gamma_0$ of the proton over a variety of focusing $n$ values. The solid line represents the predicted values; the points are the results of tracking. The simulation used $\theta_0 = 1.0$mrad, an RF-cavity and vertical focusing.}
\end{figure}

Figure \ref{mrwf1} and Figure \ref{mrwf2} shows the comparison of the precision tracking results with the analytical estimates for $\left< \Delta \gamma/\gamma_0 \right>$ and $\left< \Delta r/r_0 \right>$, respectively. (There is a vertical spread of less than one part per billion for $\left< \Delta \gamma/\gamma_0 \right>$ and less than 0.1 ppm for $\left< \Delta r/r_0 \right>$, which we assign as the error of the method.)

The results from tracking in Figures \ref{drernf}-\ref{mrwf2} match the predicted $\left<\Delta\gamma/\gamma_0\right>$ and $\left<\Delta r/r_0\right>$ values to sub-ppm accuracy. From these comparisons, we see again that the analytical estimates and the Runge-Kutta/Predictor-Corrector method pass the electric field benchmarks.

\section{Radio Frequency Wien Filter}
A radio frequency Wien Filter (WF) is a velocity-dependent charged particle filtering device. A WF can be used in a storage ring to measure a particle's electric dipole moment (EDM). The analytical estimates\citep{wien} of the EDM signal and systematic error for a particle of charge $e$, mass $m$, and anomalous magnetic moment $a$ in electric and magnetic fields are:
\[\left(\frac{ds_V}{dt}\right)_{edm} = \eta\frac{e b_V}{4 m c}\frac{(1 + a)}{\gamma^2}s_{L0}\frac{e(-E_R + \beta B_V)}{mc\omega_{a,0}},\label{edm}\]
\[\left(\frac{ds_V}{dt}\right)_{sys} =  \frac{e b_{R0}}{2 m c}\frac{(1 + a)}{\gamma^2}s_{L0},\label{sys}\]
where $s_{L0}$ is the peak longitudinal spin magnitude, $E_R$ is the radial electric field strength, and $B_V$ is the vertical magnetic field strength.  The EDM is proportional to $\eta$, with $\eta$ playing the same role for the EDM as the $g$-factor plays for the magnetic dipole moment.
The radio frequency WF ideally produces a vertical magnetic field at the $g-2$ frequency $ b_V = b_{V0} \cos{\omega_{a0} t} $ and a radial electric field $ e_R = e_{R0} \cos{\omega_{a0} t} $ with the condition\citep{wien} $ e_{R0} = \beta b_{V0}$.  However, if the WF is misaligned by an angle $\theta$ with respect to the vertical, then a radial B-field will also be present $ b_{R} = b_{R0} \cos{\omega_{a0} t} $, with $ b_{R0} = b_{V0} \sin{\theta} $ inducing a systematic error given by Equation \ref{sys}. 

A comparison between the analytical estimates and the tracking results for the deuteron case and the proton case are given in Table \ref{deut} and Table \ref{prot}, respectively.

\begin{table}[H]
\caption{Radio Frequency Wien Filter: Comparison between  analytical estimates and tracking results  for the deuteron case, in rad/s. Momentum $p$ is in GeV/c.  The EDM value is assumed to be $10^{-18}e \cdot {\rm cm}$, while for the (systematic) error, the misalignment angle is assumed to be 0.1mrad.}
\vspace{1mm}
\centering
\begin{tabular}{c c c c c} 
\hline\hline 
$p$ & EDM & EDM & Error & Error \\
&\small{Tracking} & \small{Analytic} & \small{Analytic} &\small{Tracking} \\ [0.5ex] 
\hline 
0.7  & -1.00 & -1.00 & 0.41 & 0.41 \\ 
1.4  &  -0.74 & -0.73 & 0.175 & 0.17 \\ 
2.1  &  -0.50 & -0.51 & 0.096 & 0.097 \\ 
2.8  & -0.36 & -0.35 & 0.063 & 0.06 \\ [1ex]
\hline 
\end{tabular}
\label{deut}
\end{table}

\begin{table}[H]
\caption{Radio Frequency Wien Filter: Comparison between analytical estimates and tracking results for the proton case, in rad/s. Momentum $p$ is in GeV/c.  The EDM value is assumed to be $10^{-18}e \cdot {\rm cm}$, while for the (systematic) error the misalignment angle is assumed to be 0.1mrad.}
\vspace{1mm}
\centering
\begin{tabular}{c c c c c} 
\hline\hline 
$p$ & EDM & EDM & Error & Error \\
&\small{Tracking} & \small{Analytic} & \small{Analytic} &\small{Tracking} \\ [0.5ex] 
\hline 
0.7  & 0.357 & 0.357 & 1.135 & 1.137 \\ 
1.4  &  0.174 & 0.172 & 0.393 & 0.396 \\ 
2.1  &  0.0934 & 0.093 & 0.192 & 0.195 \\ 
2.8  & 0.0566 & 0.0563 & 0.1135 & 0.1135 \\ [1ex]
\hline 
\end{tabular}
\label{prot}
\end{table}

The Wien Filter provides another benchmark for testing the accuracy of the analytical estimates and the Runge-Kutta/Predictor-Corrector tracking method. We again see very good agreement between the analytically predicted values and those calculated by tracking.

\section{Conclusion}
We have determined an array of analytical estimates for benchmarking tracking programs for precision storage ring experiments. The benchmarks form a robust test for electric and magnetic rings, with and without focusing, as well as RF cavities and Wien filters. The inclusion of a high-order contribution to the weak magnetic focusing estimate resulted in the discovery of a method to reduce or eliminate the pitch effect, which was overlooked by other authors. Together these analytical estimates give a powerful tool to benchmark programs for studying particle motion and spin dynamics in a variety of storage ring configurations.

The Runge-Kutta/Predictor-Corrector integration appears to be an accurate, albeit slow, tool for precision tracking. It has passed all the benchmarks that it was tested against, often to the part per billion level. The tracking program was able to successfully simulate particle dynamics in electric and magnetic rings with and without weak focusing, in agreement with analytical estimation. Consequently, we conclude that the Runge-Kutta/Predictor-Corrector method can be used to benchmark faster tracking programs.  Other tracking programs are also capable of providing high accuracy at least under certain conditions, e.g.,~\cite{selcuk,mane15}.  What we have presented here is an array of high precision analytical estimates that can be used to benchmark candidate simulation programs.

\section{Acknowledgements}
We would like to thank the Department of Energy and Brookhaven National Laboratory for their continued support of the High School and Supplemental Undergraduate Research Programs. We especially thank the Storage Ring EDM collaboration.  DOE partially supported this project under BNL Contract No. DE-SC0012704.  IBS-Korea partially supported this project under system code IBS-R017-D1-2014-a00.

\end{document}